\documentclass[12pt]{iopart}
\jl{1}
\eqnobysec
\usepackage{subeqn}

\def\eql{\eqalign}

\begin{document}
\jl{1}

\title{Symmetrically coupled higher-order nonlinear Schr\"{o}dinger 
equations: singularity analysis and integrability}

\author{S. Yu. Sakovich\footnote[1]{E-mail: {\tt sakovich@dragon.bas-net.by}} 
and Takayuki Tsuchida\footnote[2]{E-mail: {\tt
      tsuchida@monet.phys.s.u-tokyo.ac.jp}}
}

\address{\dag\ Institute of Physics, National Academy of Sciences, 
P.O.\ 72, Minsk, Belarus}

\address{\ddag\ Department of Physics, Graduate School of Science, 
University of Tokyo, Hongo 7-3-1, Bunkyo-ku, Tokyo 113-0033, Japan}


\begin{abstract}
The integrability of a system of two symmetrically coupled higher-order
nonlinear Schr\"{o}dinger equations with parameter coefficients 
is tested by means of the singularity
analysis. It is proven that the system passes the Painlev\'{e} test for
integrability only in ten distinct cases, of which two are new. For one of the
new cases, a Lax pair and a multi-field generalization are obtained; for the
other one, the equations of the system are uncoupled by a nonlinear transformation.\bigskip
\end{abstract}

\maketitle

\section{Introduction}

The nonlinear Schr\"{o}dinger (NLS) equation, which describes 
the time evolution of slowly varying envelope, is encountered in various 
branches of physics and known to be fundamental. In deriving the NLS 
equation as the envelope equation, we neglect higher-order terms 
under appropriate physical assumptions. 
However, due to 
recent developments in optical technology, higher-order corrections to 
the NLS equation have become necessary and important. Kodama and 
Hasegawa \cite{Kodama1,Kodama2} proposed the higher-order NLS (HONLS) equation,
\begin{equation}
q_{t}=hq_{xxx} + a q\bar{q}q_{x} + bq^{2}\bar{q}_{x} + 
\mathrm{i}(sq_{xx}+fq^{2}\bar{q})
\label{honls}
\end{equation}
which describes ultra-short pulse propagation in optical fibers, 
including higher-order effects such as higher-order dispersion, 
self-steeping and delayed Raman response. Here the bar denotes 
the complex conjugation, $\bar{q} = q^\ast$. 
For simplicity, we assume that the parameters $h, a, b, s, f$ 
are real and satisfy the 
conditions, $a^{2}+b^{2} \neq 0$, $s\neq0$ if $h=0$. 
Integrable cases of the HONLS equation \eref{honls} 
attract both theoretical and experimental interest 
because they support a variety of exact solutions and the initial-value 
problem is solvable. To date, there are four known integrable cases of the 
HONLS equation \eref{honls} which are solvable 
via the inverse scattering method:
\begin{equation}
h \neq0 \quad \; a\neq0 \quad \; b=0 \quad \; f=sa/(3h); 
\label{cs1}
\end{equation}
\begin{equation}
h \neq0 \quad \; a\neq0 \quad \; b=a/3 \quad \; f=2sa/(9h); 
\label{cs2}
\end{equation}
\begin{equation}
h = 0 \quad \; s \neq 0 \quad \; 
a\neq0 \quad \; b=0; 
\label{cs3}
\end{equation}
\begin{equation}
h = 0 \quad \; s \neq 0 \quad \; a\neq0 \quad \; b=a/2.
\label{cs4}
\end{equation}
The cases \eref{cs1}--\eref{cs4} of the HONLS equation 
\eref{honls} are called, respectively, the Hirota equation \cite{Hir}, 
the Sasa--Satsuma equation \cite{YO,SS}, 
the Chen--Lee--Liu equation \cite{CLL} 
and the Kaup--Newell equation \cite{KN}. The Painlev\'{e} analysis of the 
HONLS equation, which has been carried out by a number of authors 
\cite{Clarkson,Porse1,Mihal,Gedalin,Sako}, 
strongly indicates that eq.\ \eref{honls} is integrable only in 
the four cases \eref{cs1}--\eref{cs4}. The study based on prolongation 
structures leads to the same indication \cite{Nijhof}. 

To describe two pulses copropagating in optical fibers, we need to 
consider a two-component generalization of the single-component 
propagation equation \cite{Agrawal}. For this purpose, we consider the 
following system of two symmetrically coupled HONLS equations:
%
\begin{equation}%
\fl
\begin{array}
[c]{l}%
q_{t}=hq_{xxx}+aq\bar{q}q_{x}+bq^{2}\bar{q}_{x}+cr\bar{r}q_{x}+dq\bar{r}%
r_{x}+eqr\bar{r}_{x}+
\mathrm{i}(sq_{xx}+fq^{2}\bar{q}+gqr\bar{r})
\medskip
\\
r_{t}=hr_{xxx}+ar\bar{r}r_{x}+br^{2}\bar{r}_{x}+cq\bar{q}r_{x}+dr\bar{q}%
q_{x}+erq\bar{q}_{x}+
\mathrm{i}(sr_{xx}+fr^{2}\bar{r}+grq\bar{q})
\end{array}
\label{sys}%
\end{equation}
where $h,a,b,c,d,e,s,f,g$ are real parameters and the bar denotes the 
complex conjugation. In fact, we can assume more general conditions 
$\bar{q} = \pm q^\ast$, $\bar{r} = \pm r^\ast$ with the double signs 
being arbitrary. This makes no difference in the following analysis. 
We impose the conditions, $a^{2}+b^{2}+c^{2}+d^{2}+e^{2}\neq0$, 
$c^2+d^2+e^2+g^2 \neq 0$, $s\neq0$ if $h=0$, on the parameters so that 
the system \eref{sys} is not uncoupled and includes 
linear dispersion and higher-order nonlinearity. 
The system \eref{sys} is a natural 
generalization of the HONLS equation \eref{honls} to a two-component 
system, which is invariant under any of the transformations, 
$q \leftrightarrow r$, $q \to \e^{\mathrm{i} \alpha} q \, (\alpha : 
{\rm real})$, 
$r \to \e^{\mathrm{i} \beta} r \, (\beta: {\rm real})$. 

In this paper, we study the system of coupled HONLS equations \eref{sys} 
by means of the Painlev\'{e} analysis. Similar attempts have been 
performed for coupled NLS equations without higher-order 
terms \cite{Sahad,Park}. 
However, on the Painlev\'{e} analysis of coupled HONLS equations, only 
a few papers of academic significance have appeared \cite{Porse2,Radh}, 
where the Painlev\'{e} test has been applied to only integrable cases 
or a rather restricted class of 
equations, compared with our general form \eref{sys}. By using the Painlev\'{e} test, 
we exhaustively obtain the integrability conditions 
on the parameters 
in the coupled HONLS equations \eref{sys} for the first time. 

The paper consists of the following. 
In section~\ref{singan}, we perform the singularity analysis of \eref{sys}.
It is proven that the system \eref{sys} 
passes the Painlev\'{e} test for integrability only 
in the following ten distinct cases:%
\begin{equation}
h\neq0 \quad \; a\neq0 \quad \; b=0 \quad \; c=d=e=a \quad \; f=g=sa/(3h); 
\label{c1}%
\end{equation}
\begin{equation}
h\neq0 \quad \; a\neq0 \quad \; b=d=g=0 \quad \; c=e=a/2 \quad \; f=sa/(3h); 
\label{c2}%
\end{equation}
\begin{equation}
h\neq0 \quad \; a\neq0 \quad \; b=d=e=0 \quad \; c=a \quad \; f=g=sa/(3h); 
\label{c3}%
\end{equation}
\begin{equation}
h\neq0 \quad \; a\neq0 \quad \; b=e=0 \quad \; c=d=a/2 \quad \; f=g=sa/(3h); 
\label{c4}%
\end{equation}
\begin{equation}
h\neq0 \quad \; a\neq0 \quad \; b=d=e=a/3 \quad \; c=2a/3 \quad \; f=g=2sa/(9h); 
\label{c5}%
\end{equation}
\begin{equation}
h=0 \quad \; s\neq0 \quad \; a\neq0 \quad \; b=d=e=0 \quad \; c=a \quad \; g=f; 
\label{c6}%
\end{equation}
\begin{equation}
h=0 \quad \; s\neq0 \quad \; a\neq0 \quad \; b=c=d=e=a/2 \quad \; g=f; 
\label{c7}%
\end{equation}
\begin{equation}
h=0 \quad \; s\neq0 \quad \; a\neq0 \quad \; b=c=e=0 \quad \; d=a \quad \; g=f; 
\label{c8}%
\end{equation}
\begin{equation}
h=0 \quad \; s\neq0 \quad \; a\neq0 \quad \; b=e=a/2 \quad \; c=a \quad \; 
d=0 \quad \; g=f; \label{c9}%
\end{equation}
\begin{equation}
h=0 \quad \; s\neq0 \quad \; a\neq0 \quad \; b=d=0 \quad \; c=e=a \quad \; 
g=0. \label{c10}%
\end{equation}
In section~\ref{integr}, we show that the system \eref{sys} is integrable 
in the Lax sense in all the cases \eref{c1}--\eref{c10}. 
The integrability of 
\eref{c3}--\eref{c9} has already been studied in the literature. 
The case \eref{c2}
turns out to be related to \eref{c4}. For the case \eref{c1}, we construct a
corresponding $4\times4$\ Lax pair and propose a multi-field generalization.
For the case \eref{c10}, we obtain a nonlinear transformation, which changes
the system \eref{sys} into two independent Chen--Lee--Liu equations. 
The last section, section~\ref{conclu}, is devoted to concluding remarks.

\section{Singularity analysis\label{singan}}

\subsection{Preliminaries}

Let us apply the Painlev\'{e} test for integrability to the system
\eref{sys}. It would be better to say ``the Kovalevskaya test'' \cite{Kov},
because we, following the Weiss--Kruskal algorithm of the singularity analysis
\cite{WTC,JKM}, select only those cases of the tested system, in which
its general solution admits series expansions of Laurent type.

With respect to $q,\bar{q},r,\bar{r}$, which should be considered as mutually
independent, the system \eref{sys} is a normal system of four PDEs, of total
order twelve if $h\neq0$ or eight if $h=0,s\neq0$. A hypersurface
$\phi(x,t)=0$ is non-characteristic for \eref{sys} if $\phi_{x}\neq0$, and we
set $\phi_{x}=1$. The substitution of the expansions
\begin{equation}%
\begin{array}
[c]{l}%
q=q_{0}(t)\phi^{\alpha}+\cdots+q_{n}(t)\phi^{n+\alpha}+\cdots
\\
\bar{q}=\bar{q}_{0}(t)\phi^{\beta}+\cdots+\bar{q}_{n}(t)\phi^{n+\beta}%
+\cdots
\\
r=r_{0}(t)\phi^{\gamma}+\cdots+r_{n}(t)\phi^{n+\gamma}+\cdots
\\
\bar{r}=\bar{r}_{0}(t)\phi^{\delta}+\cdots+\bar{r}_{n}(t)\phi^{n+\delta}+\cdots
\end{array}
\label{exp}%
\end{equation}
(the bar does not mean the complex conjugation now) into the system
\eref{sys} determines the branches, i.e.\ admissible choices of $\alpha
,\beta,\gamma,\delta$ and $q_{0},\bar{q}_{0},r_{0},\bar{r}_{0}$, as well as
the positions $n$ of resonances for those branches.

We require that the system \eref{sys} admits a singular generic branch, where
the exponents $\alpha,\beta,\gamma,\delta$ are integer and at least one of
them is negative, the number of resonances is equal to the total order of the
system, all the resonances but one lie in nonnegative integer positions, and
the recursion relations for the coefficients of \eref{exp} are consistent at
the resonances. The singular generic branches are studied in
sections~\ref{sg_h_1} and~\ref{sg_h_0} for $h\neq0$ and $h=0$, respectively;
other branches are concisely 
considered in section~\ref{nongen}. Computations are done
by means of the \textit{Mathematica} system \cite{Wol}, and we omit
inessential details for this reason.

\subsection{Singular generic branches: 
$ h\neq0 $\label{sg_h_1}}

When $h \neq 0$, we set $h=1$ without loss of generality. 
The consideration of dominant terms
of \eref{sys} leads us to the following two cases to be studied separately:
$\alpha+\beta=\gamma+\delta=-2$ and $\alpha+\beta=-2,\gamma+\delta>-2$ (for
the reason of symmetry, we omit $\alpha+\beta>-2,\gamma+\delta=-2$).

\subsubsection{Case $ \alpha+\beta=\gamma+\delta=-2 $}

In this case, we have two possibilities: either two or three of the resonances
lie in the position $n=0$.

\paragraph{\bf Two resonances at {\boldmath$n=0$}.}

If we set $\alpha\neq\beta$ or $\gamma\neq\delta$, then the positions of four
resonances are $n=-1,0,0,4$, and there are 54 distinct cases of admissible
positions for other eight resonances. In none of the 54 cases, however,
$\alpha,\beta,\gamma,\delta$ are integer simultaneously.

Therefore we choose $\alpha=\beta=\gamma=\delta=-1$. Then we find from
\eref{sys} and \eref{exp} that $q_{0}\bar{q}_{0}=r_{0}\bar{r}_{0}%
=\mathrm{constant}\neq0$ (we set $\mathrm{constant}=1$ w.l.g.), and that
\begin{equation}
a=-6-b-c-d-e
\label{c1a}%
\end{equation}
\begin{equation}%
\fl
\begin{array}
[c]{c}%
(n+1)n^{2}(n-3)(n-4)
(n^{2}-6n-2b-2d+5)(n^{2}-6n-2b-2e+5)\times\\
(n^{3}-6n^{2}+(5-2d-2e)n+4(c+d+e+3))=0.
\end{array}
\label{c1n}%
\end{equation}
Five resonances lie in the positions $n=-1,0,0,3,4$, and, denoting the
positions of other seven resonances as $n_{1},n_{2},\ldots,n_{7}$, we find
from \eref{c1n} that
\begin{equation}%
\begin{array}
[c]{l}%
n_{2}=6-n_{1} \qquad n_{4}=6-n_{3} \qquad n_{7}=6-n_{5}-n_{6} \smallskip\\
d=\frac{1}{2}(5-2b-6n_{1}+n_{1}^{2}) \qquad e=\frac{1}{2}(5-2b-6n_{3}+n_{3}%
^{2}) \smallskip\\
b=\frac{1}{4}(5-6n_{1}+n_{1}^{2}-6n_{3}+n_{3}^{2}+6n_{5}-n_{5}^{2}%
+6n_{6}-n_{6}^{2}-n_{5}n_{6}) \smallskip\\
c=\frac{1}{4}(-22+12n_{5}-2n_{5}^{2}+12n_{6}-2n_{6}^{2}-8n_{5}n_{6}+n_{5}%
^{2}n_{6}+n_{5}n_{6}^{2}).
\end{array}
\label{c1r}%
\end{equation}
Taking into account the admissible multiplicity of the resonances, we have 23
distinct cases of their positions. In 22 of those cases, however, the
recursion relations for the coefficients of \eref{exp} turn out to be
inconsistent at the resonances (more details can be found in \cite{ST1}). The
only good case is $n_{1}=2,n_{3}=2,n_{5}=1,n_{6}=2$, when
$n=-1,0,0,1,2,2,2,3,3,4,4,4$, $a=c=d=e=-3/2$ and $b=0$ due to \eref{c1r} and
\eref{c1a}. In this case, we have to set $f=g=-s/2$ for the recursion
relations to become consistent at $n=2,3$. Up to a scale transformation of
variables, this is the case \eref{c1} of the system \eref{sys}.

\paragraph{\bf Three resonances at {\boldmath$n=0$}.}

In this case, resonances can lie in admissible positions only if $q_{0}\bar
{q}_{0}+r_{0}\bar{r}_{0}=\mathrm{constant}$ (we set $\mathrm{constant}=1$
w.l.g.) and%
\[%
\fl
\alpha=\beta=\gamma=\delta=-1  \qquad a=-6-b \qquad c=-6-d-e 
\qquad (b-d)(b-e)=0.
\]

If $b=d,e\neq d$, we have%
\[%
(n+1)n^{3}(n-3)(n-4)(n^{2}-6n+5-2d)
(n^{2}-6n+5-d-e)^{2}=0
\]
i.e.\ six resonances lie in the positions $n=-1,0,0,0,3,4$. Denoting the
positions of other six resonances as $n_{1},n_{2},\ldots,n_{6}$, we obtain%
\[%
\begin{array}
[c]{l}%
n_{2}=6-n_{1} \qquad n_{3}\neq n_{1} \qquad n_{4}=6-n_{3} \qquad 
n_{5}=n_{3} \qquad n_{6}=n_{4} \smallskip\\
d=\frac{1}{2}(5-6n_{1}+n_{1}^{2}) \qquad e=5-d-6n_{3}+n_{3}^{2}.
\end{array}
\]
There are four admissible choices of $n_{1},n_{3}$. The cases $n_{1}%
=2,n_{3}=1$, $n_{1}=3,n_{3}=1$ and $n_{1}=3,n_{3}=2$ lead to inconsistent
recursion relations for the coefficients of \eref{exp}. In the case
$n_{1}=1,n_{3}=2$, when $a=-6,b=d=0,c=e=-3$, we have to set $f=-2s,g=0$ for
the recursion relations to become consistent. Up to a scale transformation of
variables, this is the case \eref{c2} of \eref{sys}.

If $b=e$, we have%
\[%
(n+1)n^{3}(n-3)(n-4)(n^{2}-6n+5-2e)
(n^{2}-6n+5-d-e)^{2}=0
\]
i.e.\ six resonances lie in the positions $n=-1,0,0,0,3,4$. Denoting the
positions of other six resonances as $n_{1},n_{2},\ldots,n_{6}$, we obtain%
\[%
\begin{array}
[c]{l}%
n_{2}=6-n_{1} \qquad n_{4}=6-n_{3} \qquad n_{5}=n_{3} \qquad 
n_{6}=n_{4} \smallskip\\
e=\frac{1}{2}(5-6n_{1}+n_{1}^{2}) \qquad d=5-e-6n_{3}+n_{3}^{2}.
\end{array}
\]
There are six admissible choices of $n_{1},n_{3}$. The cases $n_{1}=2,n_{3}%
=1$, $n_{1}=3,n_{3}=1$ and $n_{1}=3,n_{3}=2$ lead to inconsistent recursion
relations for the coefficients of \eref{exp}. In the case $n_{1}=1,n_{3}=1$,
when $a=c=-6,b=d=e=0$, we have to set $f=g=-2s$ to make the recursion
relations consistent at the resonances, thus obtaining the case \eref{c3} of
\eref{sys}. The case $n_{1}=1,n_{3}=2$ with $a=-6,b=e=0,c=d=-3$, where we
have to set $f=g=-2s$, leads us to the case \eref{c4} of \eref{sys}. The
case $n_{1}=2,n_{3}=2$ with $a=-9/2,b=d=e=-3/2,c=-3$, where we have to set
$f=g=-s$, leads us to the case \eref{c5} of \eref{sys}.

\subsubsection{Case $\alpha+\beta=-2,\gamma+\delta>-2$}

We set $q_{0}\bar{q}_{0}=1$ without loss of generality. If $\alpha\neq\beta$
or $\gamma\neq\delta$, there are two or more resonances in negative positions.
Therefore we set $\alpha=\beta=-1,\gamma=\delta>-1$ and obtain%
\begin{equation}%
\begin{array}
[c]{l}%
a=-6-b \qquad d=-e+(2+c)\delta-3\delta^{2}+\delta^{3} \smallskip\\
(n+1)n^{3}(n-3)(n-4)(n^{2}-6n+5-2b)\times\smallskip\\
\left( n^{2}-3\left(  1-\delta\right)  n+2+c-6\delta+3\delta^{2}\right)^{2}=0
\end{array}
\label{nco}%
\end{equation}
i.e.\ six resonances lie in the positions $n=-1,0,0,0,3,4$. Denoting the
positions of other six resonances as $n_{1},n_{2},\ldots,n_{6}$, we find from
\eref{nco} that $n_{2}=6-n_{1}$, $n_{1}=1,2,3$, $b=\frac{1}{2}(5-6n_{1}%
+n_{1}^{2})$, $\delta=0$, $n_{3}=n_{5}=1$, $n_{4}=n_{6}=2$, $c=0$ and $d=-e$.
Then we check the consistency of recursion relations at the resonances. In the
case $n_{1}=1$, we have to set $e=g=0$ and $f=-2s$ at $n=1$, and the system
\eref{sys} becomes two independent Hirota equations \cite{Hir}. In the case
$n_{1}=2$, we have to set $e=g=0$ at $n=1$ and $f=-s$ at $n=3$, thus obtaining
two independent Sasa--Satsuma equations \cite{YO,SS}. 
In the case $n_{1}=3$, the
recursion relations are inconsistent at $n=3$.

\subsection{Singular generic branches: $h=0$\label{sg_h_0}}

When $h = 0$, we set $s=1$ without loss of generality. 
The consideration of dominant terms
of \eref{sys} leads us to the following two cases to be studied separately:
$\alpha+\beta=\gamma+\delta=-1$ and $\alpha+\beta=-1,\gamma+\delta>-1$ (for
the reason of symmetry, we omit $\alpha+\beta>-1,\gamma+\delta=-1$).

\subsubsection{Case $\alpha+\beta=\gamma+\delta=-1$}

In this case, we have two possibilities: either two or three of the resonances
lie in the position $n=0$.

\paragraph{\bf Two resonances at {\boldmath$n=0$}.}

If both $q_{0}\bar{q}_{0}$ and $r_{0}\bar{r}_{0}$ are some fixed nonzero
constants, then the recursion relations turn out to be inconsistent at the
resonance $n=1$.

\paragraph{\bf Three resonances at {\boldmath$n=0$}.}

We set $q_{0}\bar{q}_{0}=\mathrm{i}+\epsilon r_{0}\bar{r}_{0}$ w.l.g., and
then find two possibilities: $\epsilon=-1$ and $\epsilon=1$.

When $\epsilon=-1$, we have%
\[%
\begin{array}
[c]{l}%
a=\frac{-2-5\alpha-5\alpha^{2}}{1+2\alpha} \qquad 
b=e=\frac{-3\alpha-3\alpha^{2}%
}{1+2\alpha} \qquad d=a-c \qquad \gamma=\alpha \medskip\\
(n+1)n^{3}(n-2)(n-3)(n-1-k)(n-3+k)=0
\end{array}
\]
where $k=-c-2\alpha$, and only $k=0,1,2$ are admissible. The case $k=2$ is
related to $k=0$ through $q\leftrightarrow\bar{q},r\leftrightarrow\bar
{r},x\rightarrow-x,t\rightarrow-t$. In the case $k=1$, the recursion relations
are inconsistent at the resonances $n=2$. In the case $k=0$, we have to set
$g=f$ at the resonance $n=1$, and then the recursion relations are consistent
at $n=2,3$ if and only if $c=2$, $c=-2$, $c=0$ or $c=4$, which, after scale
transformations of variables, give us the cases \eref{c6}, \eref{c7},
\eref{c8} and \eref{c9} of the system \eref{sys}, respectively.

When $\epsilon=1$, we have%
\[%
\begin{array}
[c]{l}%
a=\frac{-2-5\alpha-5\alpha^{2}}{1+2\alpha} \qquad 
b=-d=\frac{-3\alpha-3\alpha^{2}%
}{1+2\alpha} \qquad e=-a-c \qquad \gamma=-1-\alpha \medskip\\
(n+1)n^{3}(n-2)(n-3)(n-1-l)(n-3+l)=0
\end{array}
\]
where $l=c-2\alpha$. In all the admissible cases $l=0,1,2$, however, the
recursion relations are inconsistent at the resonance $n=2$.

\subsubsection{Case $\alpha+\beta=-1,\gamma+\delta>-1$}

We set $q_{0}\bar{q}_{0}=\mathrm{i}$ w.l.g., and then find the following two
possibilities to have the resonances in admissible positions (see \cite{ST2}
for more details):%
\begin{equation}
\begin{array}
[c]{l}%
a=\frac{-2-5\alpha-5\alpha^{2}}{1+2\alpha} \qquad b=\frac{-3\alpha-3\alpha^{2}%
}{1+2\alpha} \medskip
\\
c=\frac{(1+2\alpha)e+(1+\alpha)(\gamma-\gamma^{2})-\alpha(\delta-\delta^{2}%
)}{(1+\alpha)\gamma+\alpha\delta} \qquad d=\frac{(\alpha\gamma+(1+\alpha
)\delta)e+\gamma\delta(2-\gamma-\delta)}{(1+\alpha)\gamma+\alpha\delta
} \medskip
\\
(n+1)n^{3}(n-2)(n-3)(n+\gamma+\delta+\frac{(1+2\alpha)e-(2\alpha+\gamma
)\delta}{(1+\alpha)\gamma+\alpha\delta})\times\medskip\\
(n+\gamma+\delta-2-\frac{(1+2\alpha)e-(2\alpha+\gamma)\delta}{(1+\alpha
)\gamma+\alpha\delta})=0;\medskip
\end{array}
\label{t10}
\end{equation}
\begin{equation}
\begin{array}
[c]{l}%
a=\frac{-2-5\alpha-5\alpha^{2}}{1+2\alpha} \qquad b=\frac{-3\alpha-3\alpha^{2}%
}{1+2\alpha} \qquad d=e=\gamma=\delta=0 \medskip\\
(n+1)n^{3}(n-2)(n-3)(n-1+c)(n-1-c)=0.\medskip
\end{array}
\label{ncd}
\end{equation}

In the case \eref{t10}, taking into account that four of the resonances
should lie in positive integer positions and that $\gamma,\delta$ should be
integer, we have to set $e=\delta-\delta^{2}/(1+2\alpha)$ and $\gamma=-\delta
$. Now the positions of resonances are $n=-1,0,0,0,1,1,2,3$. At $n=1$, we have
to set $g=0,\delta=-1-2\alpha,\alpha(\alpha+1)=0$, and then the recursion
relations become consistent at $n=2,3$ as well. Both choices of $\alpha$,
$\alpha=0$ and $\alpha=-1$, lead to the case \eref{c10} of \eref{sys}.

In the case \eref{ncd}, we have to set $c=0$ to ensure admissible positions
of resonances. Then we obtain $g=0$ at $n=1$, and the system \eref{sys}
becomes a pair of uncoupled equations.

\subsection{Other branches\label{nongen}}

We have proven that the system \eref{sys} admits good singular generic
branches in the cases \eref{c1}--\eref{c10} only. In each of those cases,
however, the system \eref{sys} admits many other branches. They are Taylor
expansions, which all are governed by the Cauchy--Kovalevskaya theorem in the
case of \eref{sys} and need no analysis, 
as well as singular
non-generic branches, which all have to be studied. All the singular
non-generic branches of the cases \eref{c1}--\eref{c10} of \eref{sys} turn
out to be good, in the sense that the exponents $\alpha,\beta,\gamma,\delta$
and the positions $n$ of resonances are integer and the recursion relations
are consistent. Omitting here the lengthy consideration of all cases, we give
as an illustration the following two singular non-generic branches of the case
\eref{c1} (where $h=1,a=-3/2$):

\begin{enumerate}
\item $\alpha=-1$, $\beta=-1$, $\gamma=-2$, $\delta=2$, $q_{0}\bar{q}_{0}=4$,
positions of resonances are $n=-4,-1,0,0,0,1,1,3,4,4,5,5$;

\item $\alpha=-2$, $\beta=0$, $\gamma=-3$, $\delta=2$, $q_{0}\bar{q}_{0}=8$,
positions of resonances are $n=-5,-2,-1,0,0,0,2,4,5,5,6,7$.
\end{enumerate}

We can conclude now that the system \eref{sys} passes the Painlev\'{e} test
for integrability only in the cases \eref{c1}--\eref{c10}. Let us establish 
in the succeeding section that the system \eref{sys} in all the cases 
\eref{c1}--\eref{c10} possesses a Lax representation. 

\section{Integrability\label{integr}}

\subsection{Known cases and new cases}

First of all, let us note that, by the transformation%
\[%
\begin{array}
[c]{l}%
x\rightarrow x+\rho t \qquad t\rightarrow t\\
q\rightarrow q\exp(\mathrm{i}\omega) \qquad \bar{q}\rightarrow\bar{q}%
\exp(-\mathrm{i}\omega) \\
r\rightarrow r\exp(\mathrm{i}\omega) \qquad \bar{r}\rightarrow\bar{r}%
\exp(-\mathrm{i}\omega)\\
\omega=\sigma x+\tau t
\end{array}
\]
with appropriately chosen real constants $\rho,\sigma,\tau$, we can make
$s=f=g=0$ in the cases \eref{c1}--\eref{c5} of \eref{sys} and $f=g=0$ in
the cases \eref{c6}--\eref{c10} of \eref{sys}.

If we set $s=f=g=0$ in \eref{c2} and \eref{c4}, these two cases of
\eref{sys} turn out to be simply related by $q\rightarrow q,\bar
{q}\rightarrow\bar{q},r\rightarrow\bar{r},\bar{r}\rightarrow r$. Since the
case \eref{c4} is known to be integrable \cite{TP}, the case \eref{c2} is
integrable as well.

The integrability of the system \eref{sys} with \eref{c3}, written in a form
of coupled modified Korteweg--de Vries (mKdV) equations, was proven in
\cite{TW1} (see also \cite{Svi}).

A Lax pair for the case \eref{c5} of \eref{sys} was given in \cite{NPSM}.

The integrability of the system \eref{sys} with \eref{c7} was proven in
\cite{MD}.

The cases \eref{c6}, \eref{c8} and \eref{c9} are integrable due to the work
\cite{TW2,TW3}.

The remaining two cases, \eref{c1} and \eref{c10}, 
of the system \eref{sys} turn out to
be new, and we prove their integrability in sections~\ref{intc1}
and~\ref{intc10}, respectively.

\subsection{Case \eref{c1} \label{intc1}}

Without loss of generality, we set $h=-1$, $a=-6$, $s=0$ 
and study the case \eref{c1} of \eref{sys} in
the form of four coupled mKdV equations 
\begin{equation}%
\begin{array}
[c]{l}%
q_{t}+q_{xxx}+6q\bar{q}q_{x}+6(qr\bar{r})_{x}=0 \\
\bar{q}_{t}+\bar{q}_{xxx}+6q\bar{q}\bar{q}_{x}+6(\bar{q}r\bar{r})_{x}=0 \\
r_{t}+r_{xxx}+6r\bar{r}r_{x}+6(q\bar{q}r)_{x}=0 \\
\bar{r}_{t}+\bar{r}_{xxx}+6r\bar{r}\bar{r}_{x}+6(q\bar{q}\bar{r})_{x}=0.
\end{array}
\label{sc1}%
\end{equation}
The compatibility condition $U_{t}-V_{x}+UV-VU=0$ of the linear problem
$\Psi_{x}=U\Psi,\Psi_{t}=V\Psi$ becomes the system of two matrix mKdV
equations \cite{AF},%
\begin{equation}%
\begin{array}
[c]{l}%
Q_{t}+Q_{xxx}-3Q_{x}RQ-3QRQ_{x}=0\\
R_{t}+R_{xxx}-3R_{x}QR-3RQR_{x}=0
\end{array}
\label{mat}%
\end{equation}
if we take the matrices $U$ and $V$ in the following block form \cite{TW1}:%
\[
U=\mathrm{i}\zeta\left(
\begin{array}
[c]{cc}%
-I_{1} & 0\\
0 & I_{2}%
\end{array}
\right)  +\left(
\begin{array}
[c]{cc}%
0 & Q\\
R & 0
\end{array}
\right)  
\]%
\begin{eqnarray}
V &=& \mathrm{i}\zeta^{3}\left(
\begin{array}
[c]{cc}%
-4I_{1} & 0\\
0 & 4I_{2}%
\end{array}
\right)  +\zeta^{2}\left(
\begin{array}
[c]{cc}%
0 & 4Q\\
4R & 0
\end{array}
\right)  +
\mathrm{i}\zeta\left(
\begin{array}
[c]{cc}%
-2QR & 2Q_{x}\\
-2R_{x} & 2RQ
\end{array}
\right)  +
\nonumber \\
&&
\left(
\begin{array}
[c]{cc}%
Q_{x}R-QR_{x} & -Q_{xx}+2QRQ\\
-R_{xx}+2RQR & R_{x}Q-RQ_{x}%
\end{array}
\right)  
\nonumber
\end{eqnarray}
where $I_{1}$ and $I_{2}$ are unit matrices, $\zeta$ is a parameter. The
choice%
\[
Q=\left(
\begin{array}
[c]{cc}%
q & r\\
\bar{r} & \bar{q}%
\end{array}
\right)  
\qquad 
R=-\left(
\begin{array}
[c]{cc}%
\bar{q} & r\\
\bar{r} & q
\end{array}
\right)
\]
changes the system \eref{mat} into \eref{sc1}. This proves that the new case
\eref{c1} of the coupled 
HONLS equations
\eref{sys} possesses a parametric Lax pair. The existence of a Lax pair 
indicates that the system \eref{sc1} 
is solvable under appropriate boundary conditions via the inverse 
scattering method. Indeed, we can solve the initial-value problem 
and obtain the $N$-soliton solution of \eref{sc1} 
by imposing appropriate constraints on the matrix-valued scattering data 
\cite{TW1}. 

Usually, it is straightforward to obtain a multi-field extension 
of integrable two-component 
systems based on the Lax formulation. In fact, we can obtain an 
$m$-component generalization of the two-component system \eref{sys} in 
the cases \eref{c2}, \eref{c4} and \eref{c6}--\eref{c9} by 
replacing two-component vectors in the Lax pair with $m$-component 
vectors. For the case \eref{c5}, we have only to replace four-component 
vectors in the Lax pair with $2m$-component vectors (this is clear if 
we express the Sasa--Satsuma equation in the form of coupled mKdV 
equations \cite{YO}). However, a multi-field generalization of 
the case \eref{c1} as well as the case \eref{c3} is not straightforward. 
Thus we explain a multi-component generalization of the system \eref{sc1},  
which is analogous to that of the case \eref{c3} (cf \cite{TW1}). 
%
For this purpose, we choose the matrices $Q$ and $R$ as
\[%
\eql{
Q=\left(
\begin{array}
[c]{cc}%
u_{0}I\otimes I+\sum_{k=1}^{2m-1}u_{k}e_{k}\otimes I & v_{0}I\otimes
I+\sum_{k=1}^{2m-1}v_{k}I\otimes e_{k}\\
v_{0}I\otimes I-\sum_{k=1}^{2m-1}v_{k}I\otimes e_{k} & u_{0}I\otimes
I-\sum_{k=1}^{2m-1}u_{k}e_{k}\otimes I
\end{array}
\right)  \smallskip\\
R=-\left(
\begin{array}
[c]{cc}%
u_{0}I\otimes I-\sum_{k=1}^{2m-1}u_{k}e_{k}\otimes I & v_{0}I\otimes
I+\sum_{k=1}^{2m-1}v_{k}I\otimes e_{k}\\
v_{0}I\otimes I-\sum_{k=1}^{2m-1}v_{k}I\otimes e_{k} & u_{0}I\otimes
I+\sum_{k=1}^{2m-1}u_{k}e_{k}\otimes I
\end{array}
\right)  
}
\]
where $I$ is the $2^{m-1}\times2^{m-1}$ unit matrix, and $\{e_{1}%
,\ldots,e_{2m-1}\}$ are $2^{m-1}\times2^{m-1}$ anti-commutative and
anti-Hermitian matrices:%
\[
\{e_{i},e_{j}\}_{+}=-2\delta_{ij}I \qquad e_{k}^{\, \dagger}=-e_{k}.
\]
Then the matrix mKdV equations \eref{mat} become the system%
\[%
\eql{
u_{j,t}+u_{j,xxx}+6\sum_{k=0}^{2m-1}u_{k}^{2}u_{j,x}+6\left(  \sum
_{k=0}^{2m-1}v_{k}^{2}u_{j}\right)  _{x}=0 \smallskip\\
v_{j,t}+v_{j,xxx}+6\sum_{k=0}^{2m-1}v_{k}^{2}v_{j,x}+6\left(  \sum
_{k=0}^{2m-1}u_{k}^{2}v_{j}\right)  _{x}=0 \smallskip
}
\qquad
j=0,1,\ldots,2m-1.
\]
Assuming that $u_{k}$ and $v_{k}$ are real and setting%
\[%
\begin{array}
[c]{c}%
u_{2j-2}+\mathrm{i}u_{2j-1}=q_{j} \\
v_{2j-2}+\mathrm{i}v_{2j-1}=r_{j}
\end{array}
\qquad
j=1,2,\ldots,m
\]
we obtain a multi-component generalization of \eref{sc1}:%
\[%
\eql{
q_{j,t}+q_{j,xxx}+6\sum_{k=1}^{m}\left|  q_{k}\right|  ^{2}q_{j,x}+6\left(
\sum_{k=1}^{m}\left|  r_{k}\right|  ^{2}q_{j}\right)  _{x}=0 \smallskip\\
r_{j,t}+r_{j,xxx}+6\sum_{k=1}^{m}\left|  r_{k}\right|  ^{2}r_{j,x}+6\left(
\sum_{k=1}^{m}\left|  q_{k}\right|  ^{2}r_{j}\right)  _{x}=0 \smallskip
}
\qquad j=1,2,\ldots,m.
\]

\subsection{Case \eref{c10}\label{intc10}}

Without loss of generality, we set $a=2$, $s=1$, $f=0$ and 
study the case \eref{c10} of \eref{sys}
in the form%
\begin{equation}%
\begin{array}
[c]{l}%
q_{t}=\mathrm{i}q_{xx}+2(q\bar{q}q_{x}+r\bar{r}q_{x}+qr\bar{r}_{x}%
) \smallskip\\
r_{t}=\mathrm{i}r_{xx}+2(r\bar{r}r_{x}+q\bar{q}r_{x}+rq\bar{q}_{x}).
\end{array}
\label{sc10}%
\end{equation}
Using the \textit{Mathematica} package \textbf{condens.m} \cite{GH}, we can
check that the system \eref{sc10} has two conservation laws for each rank
from 1 to (at least) 4. The fact, that the conservation laws appear by pairs,
is highly suggestive that equations \eref{sc10} can be uncoupled by
some transformation. 
Using the first conservation laws of \eref{sc10},%
\[%
\begin{array}
[c]{l}%
(q\bar{q})_{t}=\{\mathrm{i}(q_{x}\bar{q}-q\bar{q}_{x})+(q\bar{q})^{2}%
+2q\bar{q}r\bar{r}\}_{x} \smallskip\\
(r\bar{r})_{t}=\{\mathrm{i}(r_{x}\bar{r}-r\bar{r}_{x})+(r\bar{r})^{2}%
+2r\bar{r}q\bar{q}\}_{x}
\end{array}
\]
and introducing the new dependent variables by%
\begin{equation}
q=u\exp\left( \mathrm{i}\int_{x_{0}}^{x}v\bar{v}\mathrm{d}x^{\prime} \right) 
\qquad 
r=v\exp\left( \mathrm{i}\int_{x_{0}}^{x}u\bar{u}\mathrm{d}x^{\prime} \right)
\label{qruv}
\end{equation}
we find that the system \eref{sc10} is equivalent to two independent
Chen--Lee--Liu equations \cite{CLL}:%
\[
u_{t}=\mathrm{i}u_{xx}+2u\bar{u}u_{x} 
\qquad 
v_{t}=\mathrm{i}v_{xx}+2v\bar
{v}v_{x}%
\]
(here we have assumed that the dependent variables approach zero as
$x\rightarrow x_{0}$). 
Since a Lax pair for the Chen--Lee--Liu equation is well-known, 
we can straightforwardly obtain a Lax pair with two spectral 
parameters for the system \eref{sc10} by using a 
gauge transformation. 
This proves the integrability of the new case \eref{c10} of \eref{sys}. 
It is remarkable that, due to \eref{qruv}, the relations 
$|q|=|u|$, $|r|=|v|$ are satisfied. These relations show that 
solitons observed 
in channel $q$ and solitons observed 
in channel $r$ interact only in their phase parts. 

\section{Concluding remarks\label{conclu}}

In this paper, we have carried out the singularity analysis of 
a system of two symmetrically
coupled higher-order nonlinear Schr\"{o}dinger (HONLS) equations 
with sufficient arbitrariness of the coefficients. Apparently, the 
form of coupled HONLS equations \eref{sys} is more general than 
those which have been studied by means of the Painlev\'{e} 
analysis \cite{Porse2,Radh}. Assuming the general form of coupled 
HONLS equations, we have exhaustively selected as many as ten cases
in which the system passes the Painlev\'{e} test for integrability well. 
The integrability of eight cases in the Lax sense has been already known, 
while the remaining two cases turned out to be new. 
We have verified the integrability of the two 
cases by considering a reduction of integrable matrix equations and a 
nonlinear transformation into uncoupled integrable equations, 
respectively. The work developed in this paper 
corresponds to the spirit of the work \cite{Kov} by
Sofia Kovalevskaya, who originated the singularity analysis of nonlinear
physical systems. 

We should comment on the result 
obtained in this paper. First, though the form of coupled 
HONLS equations \eref{sys} is general enough within limitations of 
computer's performance, we may obtain 
new integrable systems of coupled HONLS equations by assuming a more 
general form. Secondly, comparing the integrable cases \eref{cs1}--\eref{cs4} 
of the single-component system with the integrable cases \eref{c1}--\eref{c10} 
of the two-component system, we notice that integrable one-component 
equations often have plural multi-component generalizations. 
We cannot identify each of \eref{c1}--\eref{c10} with 
a multi-component generalization of only one of \eref{cs1}--\eref{cs4}. 
For instance, the case \eref{c1} admits the reduction $r \to 0$ to the 
Hirota case \eref{cs1} as well as the reduction $r \to q$ to the 
Sasa--Satsuma case \eref{cs2}. Thirdly, study of exact solutions, 
which contain sufficiently many parameters, 
for integrable coupled HONLS equations
has not been well-developed so far and deserve further investigation. 

\ack
S.~Yu.~S. is grateful for support to the National Fund 
for Fundamental Research of Belarus. 
T.~T. appreciates a Research Fellowship of the Japan Society for 
the Promotion of Science for Young Scientists.

\section*{References}

\end{document}